# Deriving mobility-lifetime products in halide perovskite films from spectrally- and time-resolved photoluminescence

*Ye Yuan[1], Genghua Yan[1,*], Samah Akel[1], Uwe Rau[1] and Thomas Kirchartz[1,2]\**

[1]IMD3-Photovoltaik, Forschungszentrum Jülich, 52425 Jülich, Germany

[2]Faculty of Engineering and CENIDE, University of Duisburg-Essen, Carl-Benz-Str. 199, 47057 Duisburg, Germany

*Correspondence to: ge.yan@fz-juelich.de; t.kirchartz@fz-juelich.de

**Abstract**

Lead-halide perovskites are semiconductor materials with attractive properties for photovoltaic and other optoelectronic applications. However, determining crucial electronic material parameters, such as charge-carrier mobility and lifetime, is plagued by a wide range of reported values and inconsistencies caused by interpreting and reporting data originating from different measurement techniques. In this paper, we propose a method for the simultaneous determination of mobility and lifetime using only one technique: transient photoluminescence spectroscopy. By measuring and simulating the decay of the photoluminescence intensity and the redshift of the photoluminescence peak as a function of time after the laser pulse, we extract the mobility, lifetime, and diffusion length of halide perovskite films. With a voltage-dependent steady-state photoluminescence measurement on a cell, we relate the diffusion length to the external voltage and quantify its value at the maximum power point.

**Introduction**

The mobility-lifetime product of a semiconductor is one of the most decisive properties governing its suitability as a photovoltaic absorber material.(*1-8*) However, determining mobilities, lifetimes, and the resulting diffusion lengths requires different methods that are often inconsistent(*9*) due to the different injection conditions used. Furthermore, especially in the case of mobility measurements, huge differences(*10*) can result from different types of mobilities that are probed, e.g. intra-grain vs. inter-grain mobilities or in-plane vs. out-of-plane mobilities.(*11-13*) The reported mobilities of FAPbI$_3$(*14-18*) and MAPbI$_3$(*14, 19-32*) films across various literature sources vary from $10^{-1}$ to $10^3$ cm$^2$V$^{-1}$s$^{-1}$. In the case of lifetime measurements, inconsistencies can result from strong injection-level dependence of apparent lifetimes in situations where the decay dynamics are power-law rather than exponential(*33, 34*) as well as capacitive effects in devices that can be misinterpreted as recombination lifetimes.(*35-39*) Thus, it would be desirable to develop a method that allows measuring both mobility and lifetime with the same approach and enables the determination of lifetimes as a function of injection conditions. Spectrally resolved transient photoluminescence has the potential to fulfill these requirements. Photoluminescence (PL) essentially tracks the product of the electron and hole densities as a function of time. As recombination is reducing both the electron and hole density in a semiconductor, PL has always been one of the most essential methods to study recombination in lead-halide perovskites(*19, 40-45*) or other semiconductors with sufficient luminescence emission.(*46-54*) In recent years, the rapidly advancing field of perovskite solar cells and LEDs has garnered significant research interest and investment, leading to increasing utilization of PL techniques. These techniques include steady-state PL,(*55, 56*) time-resolved PL,(*19, 57*) voltage-dependent PL,(*58*) temperature-dependent PL,(*59, 60*) fluence-dependent PL,(*55, 61*) and others. Quantifying charge-carrier transport with PL is comparatively more difficult. Because the initial photogenerated charge carrier density inside the film is a function of depth due to Lambert-Beer-type effects and because lead-halide perovskites are generally lowly doped,(*62*) diffusion of electrons and holes will homogenize the carrier profile and thereby lead to a reduction in PL intensity(*63*) as well as a redshift of the PL peak due to an increase in reabsorption. This increase in reabsorption is caused by the average depth of emission moving from the front towards the middle of the device. The phenomenon of reabsorption transforms this spatial dependence of emission sites within the emitting perovskite film into a spectral dependence of the emission, a concept that has previously been exploited for instance in Si solar cells.(*64, 65*) In the context of halide perovskites, this phenomenon is pronounced and well-visible in thick single crystals.(*66*) The effect on films is, however, rather tiny and more difficult to notice and quantify. Thus, it has only recently been employed(*11*) to quantify out-of-plane diffusion coefficients in different halide perovskite films.

Here, we combine the quantification of peak shifts with our recently developed high dynamic range transient PL spectroscopy method based on the use of intensified gated CCD cameras, where several measurements using different gain settings are superimposed. We have recently shown that the differential decay times resulting from tr-PL measurements performed at different initial fluences differ in their early time decay but converge to a fluence-independent but carrier-concentration-dependent decay time at later times. Thus, the decay time is a unique function of carrier densities – as expected for recombination of a spatially homogeneous distribution of electrons and holes – only at longer times, while at earlier times it retains a memory of the laser fluence. Here, we show that the regions where decay times retain the information on the initial fluence are identical to the parts of the decay where a peak shift in the PL spectrum is visible. Thus, these regions are dominated by changes in the number of reabsorbed photons and are therefore indicative of the times, when the electron and hole concentrations are still a function of depth within the film. Consequently, we can distinguish the part of the decay that is affected by diffusion from that affected by recombination alone.

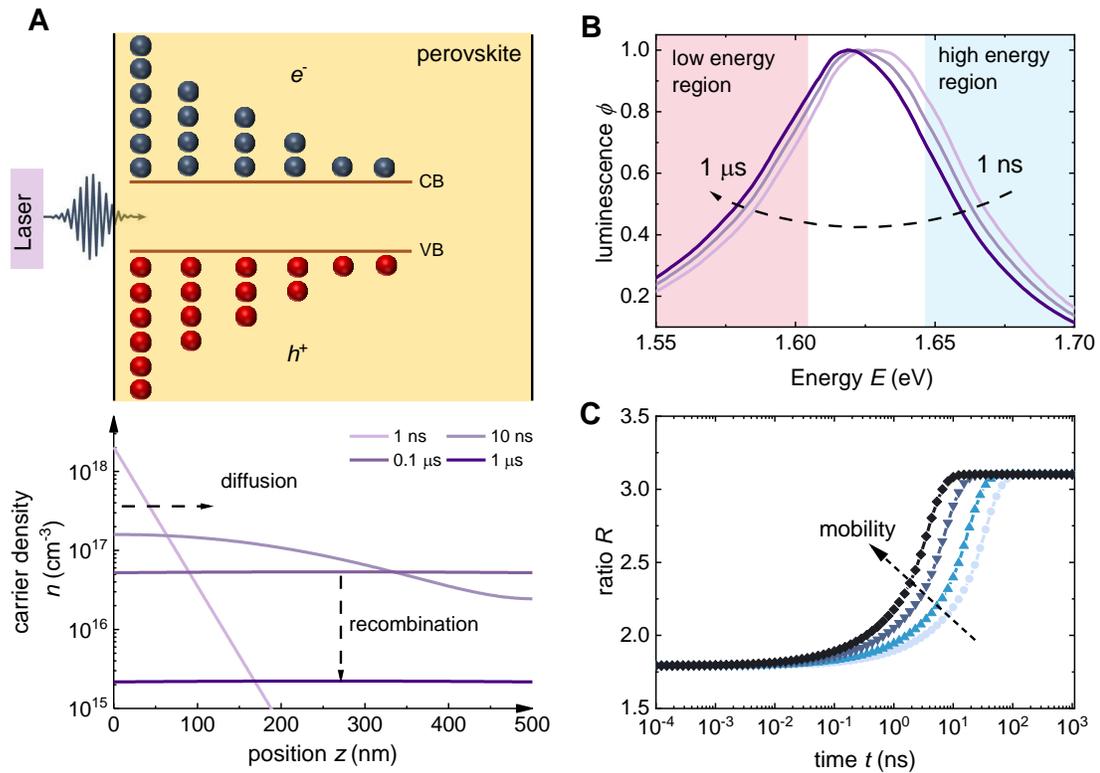

**Fig. 1. Schematic illustration of carrier diffusion and spectral shift process based on simulated results.** **(A)** After the laser pulse hits the sample, the generated carriers in the film will diffuse from one side of the film to the other, while recombination occurs simultaneously. **(B)** Time-dependent photoluminescence spectra show an obvious red shift with time, which indicates the reabsorption of photoluminescence along with carrier diffusion in perovskite. **(C)** By plotting the ratio $R$ as a function of time, we can quantify the influence of mobility and diffusion coefficients on the spectral shifts.

## Results

### Capturing Carrier Kinetics

There are essentially two options for how to detect both time-dependent and spectral information during a transient photoluminescence measurement. Option 1 is to perform two measurements with different sets of filters and then compare the two measurements. This option is easy to implement with traditional time-correlated single photon counting equipment and has recently been used by Cho et al.(*11*) Option 1 could be implemented by doing two measurements in series but it could also be done simultaneously using a beam splitter and two photodetectors. Option 2 involves using a detector, which directly captures the whole spectrum for each time delay. Here, one could use intensified CCD cameras with a variable gate or streak cameras. For the current study, we use option 2 using a gated intensified CCD camera from Andor. The investigated samples are triple-cation $Cs_{0.05}FA_{0.73}MA_{0.22}PbI_{2.56}Br_{0.44}$ perovskite films directly prepared on Corning glasses with a band gap of 1.63 eV and thickness of 550 nm.

A schematic representation of the experiment is shown in Fig. 1. After excitation, photocarriers are generated primarily close to the front side of the perovskite film. The initial carrier distribution (at point $t \approx 0$) is determined by the absorption coefficient $\alpha$ of the film (Section 2 of Supporting Information) and the wavelength $\lambda$ of the excitation laser (here, $\alpha = 4\times10^5$ cm$^{-1}$ and $\lambda = 343$ nm, respectively). Here, we intentionally use a UV laser with a low wavelength where the perovskite has a high absorption coefficient. The lower the wavelength, the higher the absorption coefficient of the perovskite layer at the laser wavelength and the more abrupt the initial carrier profile will be as a function of depth within the film. Driven by the concentration difference, the carriers diffuse from the front towards the back side of the film until the carrier concentration is approximately independent of position. Depending on the recombination velocities at the two surfaces, gradients of the concentrations of electrons and holes towards the surfaces may remain. The time it takes for the carrier concentrations to homogenize depends on the mobility of electrons and holes. In the example situation presented in Fig. 1A, where we chose a mobility of 1 cm$^2$/Vs for electrons and holes, we see that carrier diffusion mainly occurs in the first tens of nanoseconds. After 100 ns, the carrier concentrations are nearly perfectly flat and from then onwards recombination is the only driver of a further reduction in both photoluminescence and carrier densities. Using a gated CCD setup, we can obtain a series of time-dependent photoluminescence spectra, which we later transform to a transient PL decay curve. The transient decay curve provides information about charge-carrier recombination, while the time-dependent photoluminescence spectra reflect the process of carrier diffusion. The schematic illustration of the measurement technique is shown in Fig. S1. As shown in Fig. 1B, the simulated time-dependent photoluminescence spectra corresponding to the carrier profiles seen in Fig. 1A show a redshift with time caused by diffusion and photon reabsorption.(*66*) In our measurements, the detector is positioned on the same side as the excitation laser. The photoluminescence caused by radiative recombination must pass through the film before being detected and thus being reabsorbed. Reabsorption primarily reduces the luminescence on the high-energy flank of the spectrum, thereby reducing the peak energy. As the charge carriers diffuse deeper into the film, more photoluminescence is absorbed. Thus, the spectra exhibit a redshift over time.

There are different ways to quantify the spectral shift. The most intuitive option might be to plot the PL peak as a function of time after excitation. However, for noisy data, the peak itself is a poor indicator of spectral shifts as it relies on one single point of the spectrum (the highest point). A better option would be to use the whole spectrum for quantifying the shift. This could be achieved e.g. via calculating the center of mass or via determining the photoluminescence intensity ratio of the low-energy region to the high-energy region. We chose the latter approach which was already employed previously by ref. (*11*), because it will be more easily applicable to researchers using TCSPC setups for the experiment. In Fig. S2, we compare these three methods, demonstrating that the variation of the center of mass contains the same information as that of the ratio, while the change in the peak (of the experimental data) is not suitable as an effective indicator owing to its requirement for high spectral resolution and low noise. Here, we define the spectral shift ratio as $R = \int_{1.45}^{1.58} \phi \, dE / \int_{1.68}^{1.80} \phi \, dE$. The exact integration boundaries will always be somewhat arbitrary and must be adjusted to the spectral position of the PL spectrum. Thus, the absolute value of $R$ will be largely irrelevant and strongly dependent on the integration boundaries. However, the relative change in $R$ as a function of the time delay after the laser pulse will be decisive in quantifying the mobility. When the mobility increases, the ratio increases rapidly until it reaches a plateau (Fig. 1C), which indicates the equilibrium stage of the carrier distribution. In Section 2 of Supporting Information, the influence of other factors was investigated. Specifically, the absorption coefficient affects the initial value of the ratio, while the film thickness and surface recombination velocity affect the plateau value. We further investigated the influence of shallow defects, as shown in Section 3 of Supporting Information. The results demonstrate that the detrapping effect of shallow defects dramatically alters the plateau, whereas it exhibits a negligible influence in the initial 10 ns.

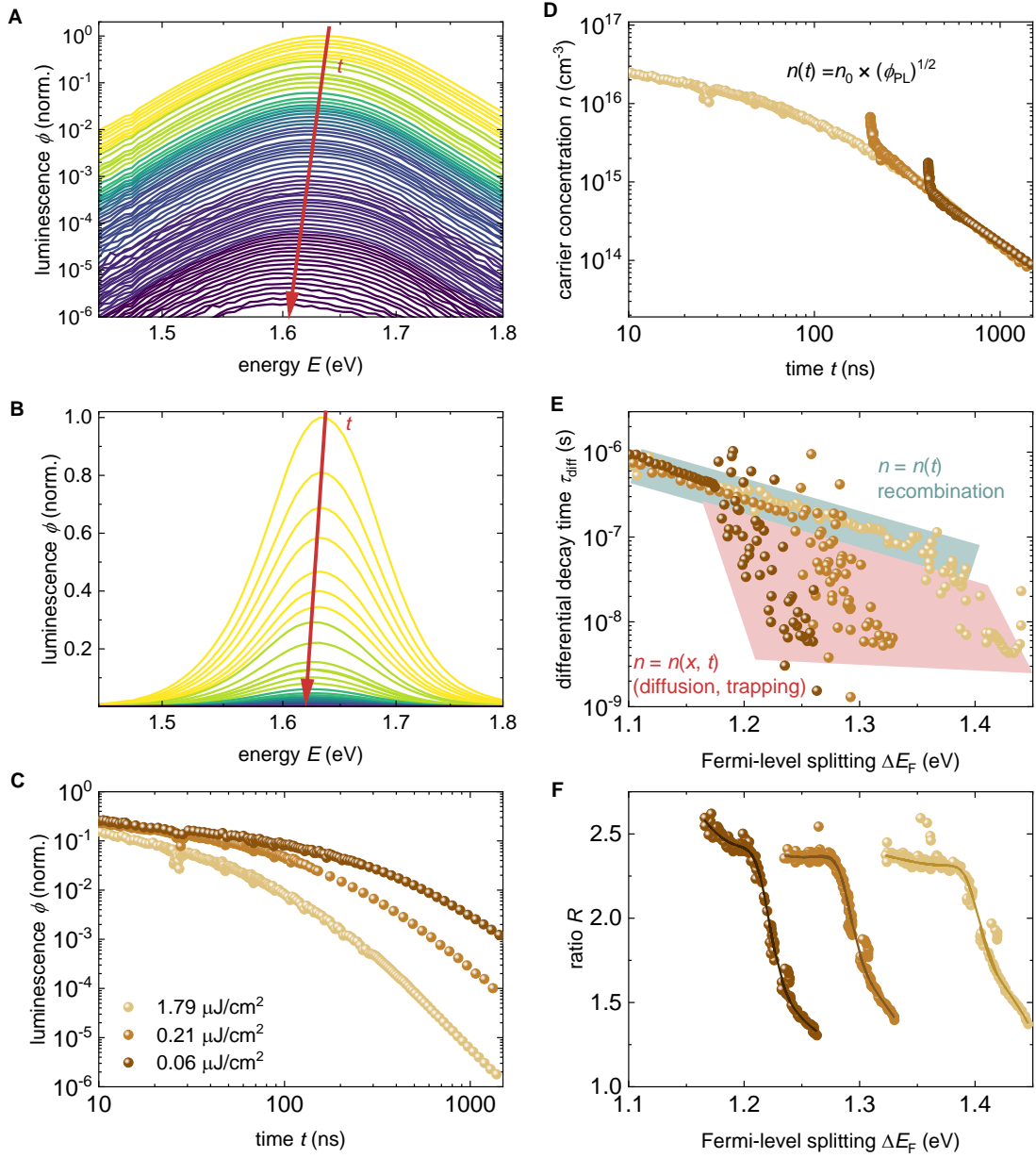

**Fig. 2. Tr-PL decay and spectral shift of experimental data for different illumination intensities.** (**A**) PL spectra at all times during the measurement with illumination intensity of 1.79 µJ/cm² using logarithmic scale. (**B**) PL spectra at all times during the measurement with an illumination intensity of 1.79 µJ/cm² using a linear scale. (**C**) Photoluminescence decay curves. (**D**) Decay curves of carrier concentration with shifted time axis. (**E**) Differential decay time vs. Fermi-level splitting. (**F**) Spectral shift ratio vs. Fermi-level splitting. The solid lines in panels (**E**) and (**F**) indicate trends in the experimental data.

**Photoluminescence Experiments**

In Fig. 2, the tr-PL data measured using the gated CCD setup with different fluences are displayed. The recorded time-dependent PL spectra in Fig. 2A exhibit an obvious peak shift. Specifically, the PL spectra at different times are shown in Fig. 2B on a linear scale which is a more frequently seen way to display PL spectra in the literature. Fig. 2C presents the tr-PL decay curves, which exhibit a slower decay at lower illumination intensities. In Fig. 2D, the luminescence is transformed into carrier concentration, and the time axis is adjusted to merge the curves. The merged curves show two distinct parts that we will refer to as the tail and the envelope. Although the envelope parts of the different curves generally overlap, the initial parts of each decay (tail part) are noticeably outside the envelope. In Fig. 2E, the $\Phi_{PL}$ vs. $t$ plot (Fig. 2C) is transformed into the $\tau_{diff}$ vs. $\Delta E_F$ plot based on $\tau_{diff} = \left(-\frac{1}{2} d\ln(\phi_{PL})/dt\right)^{-1}$, $\Delta E_F(0) = k_B T \ln(\Delta n(0)^2/n_i^2)$, and $\phi_{PL} \propto \exp(\Delta E_F/k_B T)$.(*67, 68*) The initial carrier concentration $\Delta n(0)$ after the laser pulse is estimated based on the measured laser power densities. Similar to the curves in Fig. 2D, all curves can be divided into tail and envelope parts. Additionally, the spectral shift ratio $R$ acquired from the time-dependent photoluminescence spectra is depicted in Fig. 2F. The ratio $R$ can be divided into a time-dependent and an approximately time-independent part. At early times, the carriers diffuse from the front of the film into the bulk until they are homogeneously distributed throughout the film. While the diffusion process occurs, the ratio $R$ is time dependent. Once the homogenization process is complete, the ratio remains constant. Comparing Fig. 2E and Fig. 2F, we found that the tail parts of the decay curves correspond to the time-dependent parts of the ratio $R$, indicating that carrier diffusion is one of the primary physical processes causing the "tail" phenomenon. Mathematically, we can understand the tail regime as the part of the decay that is governed by the solution of partial differential equations in time and position, i.e. $n = n(x,t)$. In contrast, the envelope part of the decay is governed by the solution of ordinary differential equations in time where we just need to track the evolution of the average carrier density $n_{av}$ vs. time, i.e. $n(x) \approx n_{av}$ and $n_{av} = n_{av}(t)$. During the tail part of the decay, it matters whether a given average carrier density was created 30 ns or 300 ns ago by the laser pulse, because the two cases will show different $n(x)$ profiles. Thus, the tail part depends on the fluence, while the envelope part only depends on the average carrier density or the Fermi-level splitting.

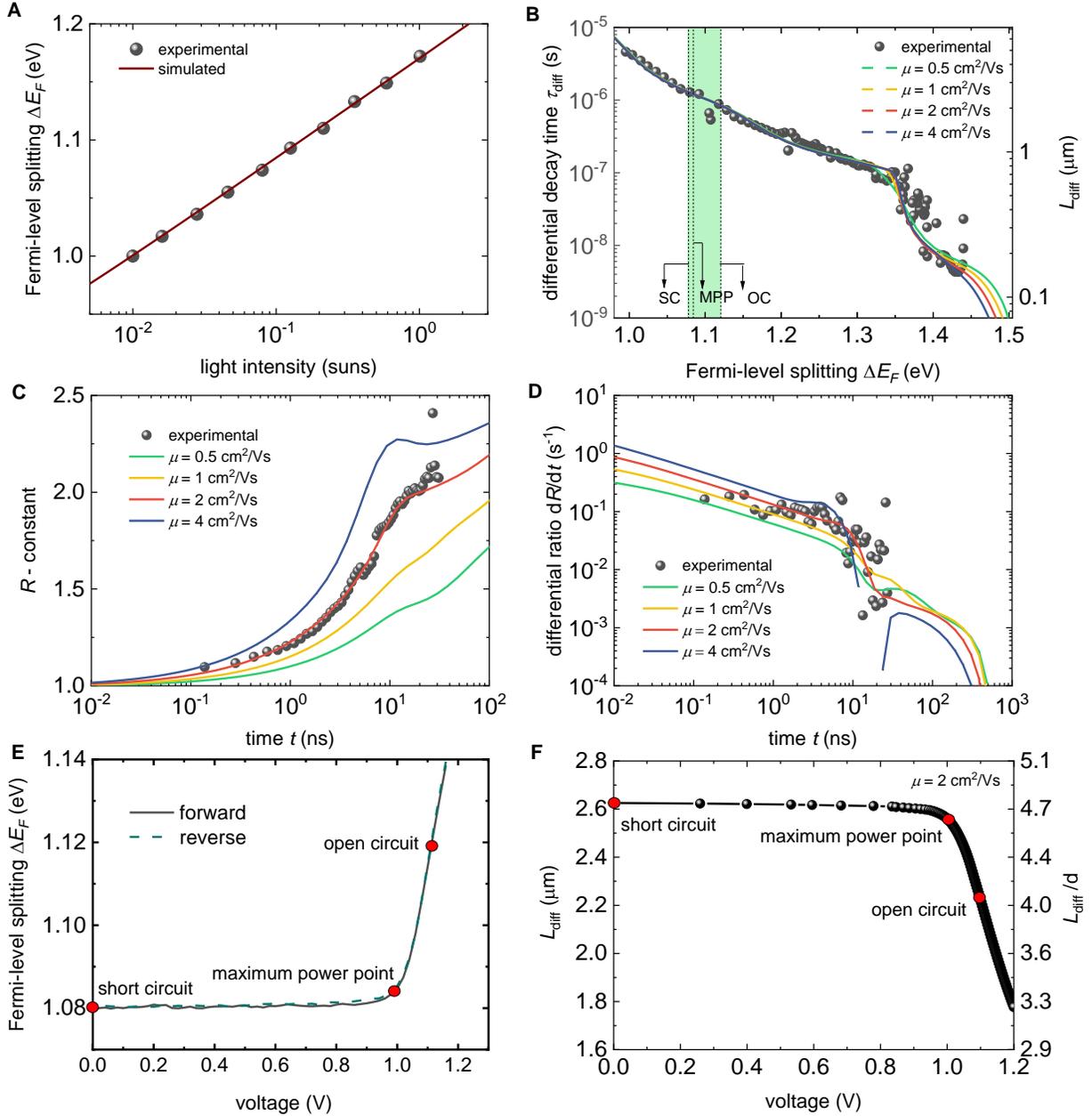

**Fig. 3. Experimental and simulation results.** (**A**) Experimental results of $\Delta E_F$ vs. illumination intensity acquired from steady-state PL results and the corresponding simulation results. (**B**) Experimental differential decay time vs. Fermi-level splitting acquired from gated CCD setup and the corresponding simulated results with different carrier mobilities. On the y-axis(right side), we calculated $L_{\text{diff}} = \sqrt{D\tau_{\text{diff}}}$, where $D = \mu\, kT/q$ and $\mu = 2$ cm$^2$/Vs. (**C**) Experimental ratio vs. time acquired from gated CCD setup and the corresponding simulated results with different carrier mobilities. Here we shift the $R$ by subtracting a constant to obtain the same starting value. (**D**) The corresponding differential ratio value vs. time from the experiment and simulations. **(E)** Experimental results of $\Delta E_F$ vs. the external bias voltage acquired from the voltage-dependent photoluminescence measurement under 1 sun light intensity. **(F)** The relationship between $L_{\text{diff}}$ and external bias voltage, which is acquired from reverse scan result of Fig. 3E and simulated result ($\mu = 2$ cm$^2$/Vs) of Fig. 3B. On the y-axis (right side), we calculated $L_{\text{diff}}/d$, where $d$ is the thickness of the film.

**Inferring the mobility by numerical simulations**

Based on the experimental steady-state and transient photoluminescence (PL) data, we perform numerical simulation to extract the mobility of our sample. The detailed description of the numerical models is found in Section 5 of the Supporting Information. To start the work, we utilized a zero-dimensional (0D) model, which primarily focus on trapping, detrapping and recombination processes, to simultaneously fit both the steady-state (Fig. 3A) and transient PL experimental data (Fig. 3B and Fig. S3), as reported in our previous work.(*34*). Fig. 3A shows that the calculated $\Delta E_F$ from steady-state PL data measured under various illumination intensities agrees well with the simulated result. Subsequently, to better understand the carrier dynamic process, we employed a one-dimensional (1D) model, which not only contains the trapping, detrapping and recombination processes but also incorporates the diffusion and reabsorption processes and is thereby able to reproduce and fit the PL peak shift. The simulated results are presented in Fig. 3B, 3C and 3D. Moreover, we compare the fitted $\tau_{\text{diff}}$ using the two models (Fig. S3) and find that the two fitted curves are almost overlapping except for the tail part, which is consistent with the fact that the 0D model does not consider the carrier diffusion process. The values of fitting parameters obtained from both models are identical, as listed in Table S1. As depicted in Fig. 3B, both the differential decay time $\tau_{\text{diff}}$ and diffusion length $L_{\text{diff}} = \sqrt{D\tau_{\text{diff}}}$ increase with decreasing $\Delta E_F$. From the simulated results for different mobilities, we observe that the impact of mobility is primarily confined to the tail parts of decay, while the envelope of the decay remained relatively unchanged. The spectral shift ratio and differential ratio are more sensitive to the variation of mobility, as demonstrated in Fig. 3C and 3D. Note that as tiny changes in bandgap energy (1 meV) would result in large differences in the value of $R$ (Fig. S4), we suggest that the variation trend of $R$ and the value of $dR/dt$ are more important for analyzing carrier diffusion. Consequently, we determined a mobility $\mu \approx 2$ cm$^2$/Vs of our perovskite film sample, as well as a diffusion coefficient $D \approx 0.052$ cm$^2$/s, which was calculated using the Einstein relation $D = \mu k_B T/q$ where $k_B T$ is the thermal energy and $q$ is the elementary charge. Note that the sample exhibits similar mobility under lower illumination intensities (Fig. S5). Fig. 3E presents the relationship between $\Delta E_F$ and the external bias voltage, which is obtained from the voltage-dependent photoluminescence measurement performed under 1 sun illumination condition. Moreover, the diffusion length as a function of the external bias voltage is displayed in Fig. 3F, which is determined by combining the results shown in Fig. 3B and Fig. 3E. We observe that the sample exhibits a $L_{\text{diff}}$ of approximately 2.6 µm at short-circuit condition and maximum power point and ~2.2 µm at open-circuit condition.

**Physical Processes Underlying the $\tau_{\text{diff}}$ vs. $\Delta E_F$ Plot**

There are three essential processes that affect the carrier concentration within a perovskite film after the excitation by a laser pulse: (I) the diffusion of carriers, (II) the trapping of free carriers by shallow defects, and (III) the transition of trapped carriers to the valence band. These processes correspond to distinct regions in the $\tau_{\text{diff}}$ vs. $\Delta E_F$ plot. In the following, we will elaborate these processes assuming for simplicity that the shallow traps are closed to the conduction band. For shallow traps close to the valence band, every instance of the word 'electron' would have to be replaced by 'hole' and vice versa.

Immediately after excitation, the photogenerated carriers begin to diffuse from the illuminated side to the other side. As demonstrated in Fig. 4A, carrier diffusion has significant impact on the onset value of $\tau_{\text{diff}}$, which in turn affects the shape of the curve in high Fermi-level region (1.43-1.5eV). Additionally, trapping by defects is another important physical process that may affect the PL decay at early times. In Fig. 4C, we adjust the electron capture coefficient $\beta_n$ of the shallowest trap, while maintaining a constant trap density $N_t$. Consequently, the electron lifetime $\tau_n$ is altered accordingly determining by $\tau_n = 1/(N_t \beta_n)$. As shown in Fig. 4C, the electron lifetime plays a crucial role in shaping the tail part. For a shorter electron lifetime, which implies more effective electron capture by the defect, the $\Delta E_F$ shows faster decrease and the trapping process is shortened. The impact of the trapping process is primarily observed in the higher $\Delta E_F$ range (e.g. 1.4-1.5 eV). Furthermore, Fig. 4C reveals that although the carrier trapping process affects the shape of the tail part, it has little influence on the onset point. In Fig. S6, a simpler scenario where only one shallow trap and a low radiative recombination coefficient is employed, further demonstrating the impact of trapping process on the tail part, which supports the aforementioned argument. The final stage of the carrier transport process involves the non-radiative recombination of trapped electrons with holes in the valence band. This process is primarily dependent on the hole lifetimes, which determines the shape of the envelope. The hole lifetime is calculated as $\tau_p = 1/(N_t \beta_p)$, where $\beta_p$ is the hole capture coefficient. As shown in Fig. 4E, we adjusted the hole lifetime $\tau_p$ for Trap 2 (the middle one) and observed that a lower hole lifetime leads to a smaller $\tau_{\text{diff}}$ for a given $\Delta E_F$. Since we included three traps in the simulation, they affect different regions of $\Delta E_F$ sequentially. From shallow to deep, the traps influence the high, middle and low $\Delta E_F$ regions, respectively (see Fig. S7 and Fig. S8).

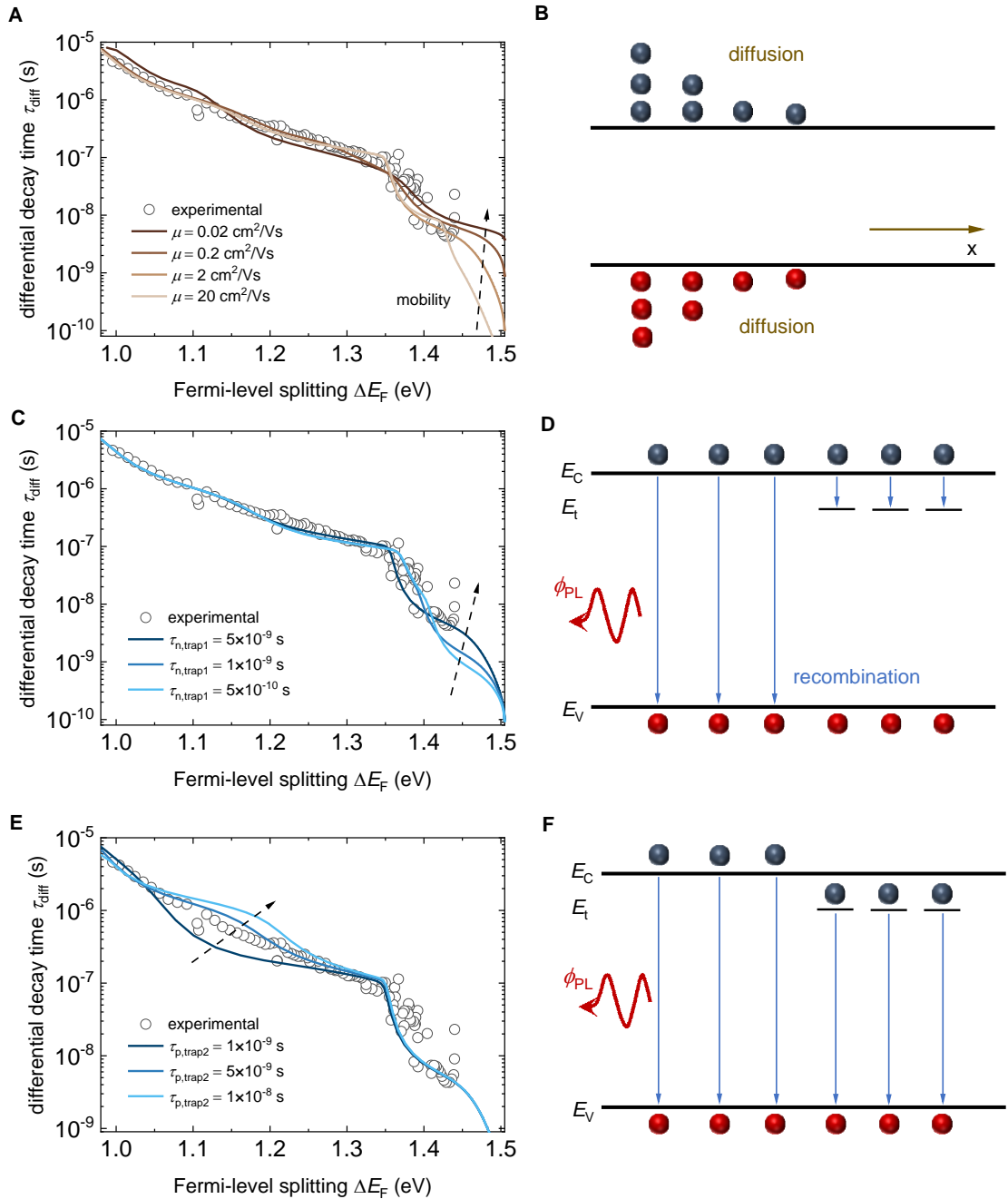

**Fig. 4. Analysis of factors influencing $\tau_{\text{diff}}$ vs. $\Delta E_F$ plot and corresponding physical processes.** (**A,B**) Influence of carrier diffusion on the onset value of $\tau_{\text{diff}}$. The carrier mobility was adjusted. (**C,D**) Influence of electron trapping process on the tail part of $\tau_{\text{diff}}$. The electron lifetime related to Trap 1 is adjusted by changing electron capture coefficient $\beta_n$, while maintaining a constant trap density $N_t$. (**E,F**) Influence of the hole capture process on the envelope part of $\tau_{\text{diff}}$. The hole lifetime related to Trap 2 was adjusted by changing hole capture coefficient $\beta_p$, while maintaining

a constant trap density $N_t$. We further show the influence of the hole lifetimes related to Trap 1 and Trap 3 in Supplementary Fig. S7 and Fig. S8.

**Numerical simulation of device performance**

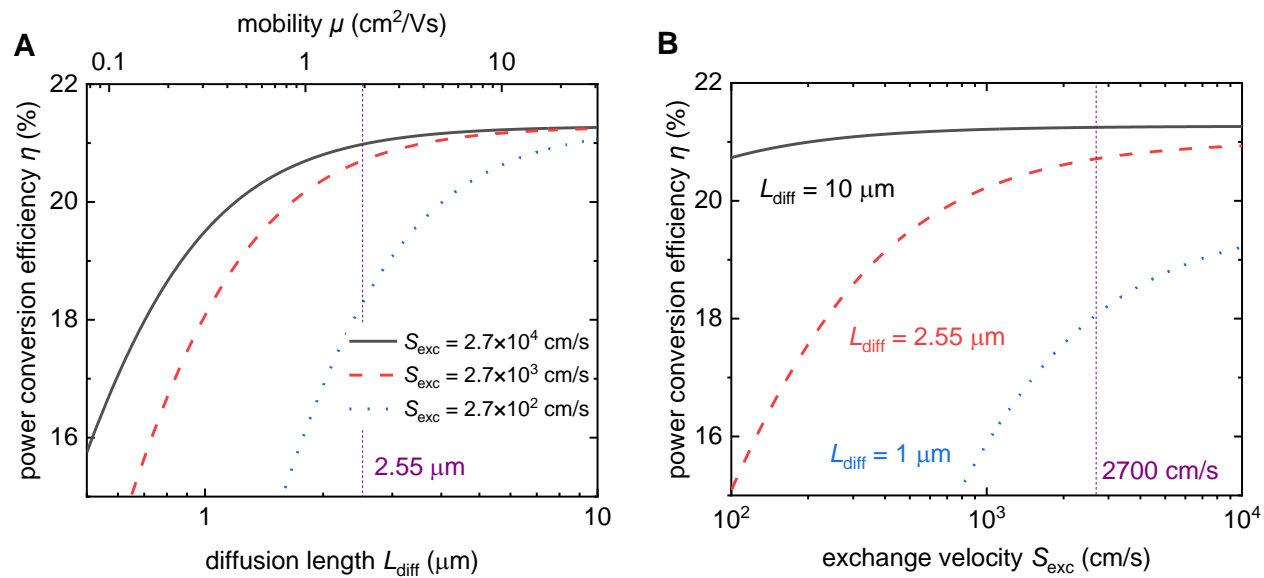

**Fig. 5 Numerical simulation of device performance.** The simulated power conversion efficiency as a function of (A) diffusion length $L_{\text{diff}}$ and (B) exchange velocity $S_{\text{exc}}$ using Equation (1). The mobility $\mu$ in (A) was calculated using $L_{\text{diff}} = \sqrt{\frac{k_B T}{q}\mu \tau_{\text{diff}}}$ by assuming $\tau_{\text{diff}}=1.26\times 10^{-6}$ s. The purple dashed lines in the figures at (A) $L_{\text{diff}}$ =2.55 µm and (B) $S_{\text{exc}}$ =2700 cm/s indicate the approximate experimental values of our sample.

Furthermore, we want to put the obtained parameters into perspective. This can be done using numerical simulations or alternatively using analytical approximations. For the sake of achieving a more intuitive approach to the influence of absorber layer diffusion and transport layer mobility on charge collection, we newly derived an approximative but analytical equation for the current-voltage curve of a perovskite solar cell. This equation effectively includes the influence of ions by assuming a perfectly field free absorber layer, where the field is screened by a sufficiently high ion density. Furthermore, it includes bulk recombination, diffusion within the perovskite and considers transport through the electron and hole transport layers by an effective exchange velocity $S_{\text{exc}}$. This exchange velocity can be easily determined from voltage-dependent PL measurements(*37*), whereas the diffusion length and bulk recombination lifetime can be deduced from the tr-PL measurements, as shown in Fig. 3. We can then write the current voltage curve as (see Section 4 of Supporting Information for detailed derivation)

$$J = qd \left[ \frac{\frac{L_{\text{diff}}}{d} \tanh\left(\frac{d}{L_{\text{diff}}}\right)}{\frac{L_{\text{diff}}}{\tau_{\text{diff}} S_{\text{exc}}} \tanh\left(\frac{d}{L_{\text{diff}}}\right) + 1} \right] \left\{ \frac{n_0}{\tau_{\text{diff}}} \left[ \exp\left(\frac{qV_{\text{ext}}}{2k_B T}\right) - 1 \right] - G \right\} \quad (1)$$

where $n_0$ is the intrinsic carrier concentration, $V_{\text{ext}}$ is the external voltage, $G$ is the average generation rate throughout the perovskite layer and $S_{\text{exc}}$ is the carrier extraction velocity of the transport layer. Our derivation of Equation (1) is based on earlier work primarily by Sandberg(*69*) and Rau(*70*) and provides a more intuitive understanding of extraction losses as compared to purely numerical simulation results. Note that Equation (1) contains recombination and transport as parameters. However, as our experimental assay of recombination originates from a film, the resulting *J-V* curve will provide a hypothetical *J-V* curve in the absence of interfacial recombination. Thus, the equation predicts a higher $V_{\text{oc}}$ than that of the actual device, whereby the difference is due to interfacial recombination losses.

The prefactor $[L_{\text{diff}} \tanh(d/L_{\text{diff}})/d]/[1 + L_{\text{diff}} \tanh(d/L_{\text{diff}})/(\tau_{\text{diff}} S_{\text{exc}})]$ in Equation (1) can be intuitively understood as a collection efficiency, which is related to $L_{\text{diff}}$ and $S_{\text{exc}}$. Note that $L_{\text{diff}}$ and $S_{\text{exc}}$ represent properties of the perovskite absorber and transport layers, respectively. The parameter $S_{\text{exc}}$ describes how fast transport of electrons through the ETL and holes through the HTL is, and at steady state, the $S_{\text{exc}}$ can be extracted from the voltage-dependent PL measurements via(*37*)

$$S_{\text{exc}} = J / \left\{ q n_0 \left[ \exp\left(\frac{qV_{\text{ext}}}{2k_B T}\right) - \exp\left(\frac{qV_{\text{int}}}{2k_B T}\right) \right] \right\} \quad (2)$$

For our device, the $S_{\text{exc}}$ is around 2700 cm/s within the range of voltages relevant for one-sun operation (Fig. S9). Other parameters for the calculation are shown in Fig. S10 and Table S2. Fig. 5A and 5B exhibit the variation of power conversion efficiency $\eta$ along with diffusion length $L_{\text{diff}}$ and exchange velocity $S_{\text{exc}}$, respectively. The efficiency was simulated by using Equation (1). As shown in Fig. 5A, with increasing of $L_{\text{diff}}$, the efficiency $\eta$ remarkably increases at the beginning, but later the increasing rate gradually slows down until it reaches a plateau. It can be observed that when the mobility increased from 1 to 10 cm$^2$/Vs (i.e. $L_{\text{diff}}$ increased from 1.81 to 5.7 μm), the efficiency of the sample with $S_{\text{exc}}$=2700 cm/s (the red curve) can increase by 4.88%. However, when the mobility dropped off by an order of magnitude (from 1 to 0.1 cm$^2$/Vs, i.e. $L_{\text{diff}}$ decreased from 1.81 to 0.57 µm), the efficiency dramatically decreased by 31.49%. Moreover, the decline would be more severe for the sample with a slower $S_{exc}$ (blue curve). The situation of the effect of $S_{exc}$ on the efficiency is similar. When the $S_{exc}$ is sufficiently fast, the increasing trend of the efficiency slows down until reaching a plateau. For a certain $L_{diff}$, for instance $L_{diff}$ =2.55 µm (the red curve in Fig. 5B), the efficiency increased by 5.14% when the $S_{exc}$ increase from $10^3$ to $10^4$ cm/s, while it dramatically decreased by 25.39% when the $S_{exc}$ dropped off from $10^3$ to $10^2$ cm/s.

Regarding Equation (1), in case of $L_{diff} \gg d$, we can make an approximation of $\tanh(d/L_{diff}) = d/L_{diff}$ by using the Taylor expansion and hence the collection efficiency can be simplified to $[d/(\tau_{diff} S_{exc}) + 1]^{-1}$. Therefore, the current-voltage curve can be written as(37)

$$J = qd \left( \frac{1}{\frac{d}{\tau_{diff} S_{exc}} + 1} \right) \left\{ \frac{n_0}{\tau_{diff}} \left[ \exp\left(\frac{qV_{ext}}{2k_B T}\right) - 1 \right] - G \right\} \quad (3)$$

The $L_{diff}$ cancels out such that the result is independent of the bulk mobility, which explains the phenomenon of the plateau for sufficiently long $L_{diff}$. In our case, the ratio $d/L_{diff} \approx 0.2$, which satisfies the approximation of $\tanh(d/L_{diff}) = d/L_{diff}$ (see Fig. S11), suggesting a collection efficiency of $[d/(\tau_{diff} S_{exc}) + 1]^{-1}$. Furthermore, in case of $\tau_{diff} S_{exc} \gg d$, the collection efficiency approaches 1 (which means sufficiently efficient charge carrier extraction); therefore, further increasing $S_{exc}$ would not have an effective impact on the efficiency. In contrast, if $S_{exc}$ is getting smaller, the denominator of the collection efficiency will rapidly increase, which explains the dramatic decrease in the efficiency with decreasing $S_{exc}$. In our case, the ratios $d/L_{diff} \approx 0.2$ and $\tau_{diff} S_{exc}/d \approx 6804$ approximately meet the conditions of $L_{diff} \gg d$ and $\tau_{diff} S_{exc} \gg d$, thereby ensuring good device performance (see Fig. S12). Therefore, while a gain in any of the parameters would lead to small gains in efficiency, a drop in $L_{diff}$ and $S_{exc}$ would have a significant negative impact on the efficiency.

**Discussion**

In this study, we investigated the carrier dynamics of perovskite thin films through a detailed analysis of spectrally- and time-resolved photoluminescence data. The data collected on both PL decay and PL redshift over time shed light on the carrier recombination and diffusion processes. Our findings indicate that the onset of time-resolved PL decay is determined by the electron lifetime with the underlying physical process being the electron capture by shallow defects. After excitation, the carriers accumulate on the excitation side, leading to carrier diffusion due to concentration differences. A "tail" part appears in the PL decay curve during this stage, while the time-dependent photoluminescence spectra exhibit a significant spectral redshift. This redshift occurs due to photon reabsorption, and gradually disappears when carriers are evenly distributed with mobility playing a dominant role in this process. The ensuing "envelope" in the decay curves is determined by the SRH recombination, which is determined by the hole lifetime of the traps. In the $\tau_{\text{diff}}$ vs. $\Delta E_\text{F}$ plot, a longer hole lifetime of the trap leads to a smaller $\tau_{\text{diff}}$ or $\Delta E_\text{F}$.

Through simulations of the PL decay and PL redshift results, we determined parameters for both carrier recombination and diffusion. Our findings suggest that the mobility and diffusion coefficient of the perovskite thin film are approximately 2 $cm^2/Vs$ and 0.052 $cm^2/s$, respectively. Additionally, we found that the lifetime increases with a lower $\Delta E_\text{F}$ value, resulting in an anti-correlation between mobility-lifetime products and injection-level. It is worth noting that we assumed acceptor-like defects close to the conduction band in our discussion. If the traps are donor-like defects close to the conduction band, the electron-hole types mentioned above would be swapped. This study provides a comprehensive understanding of the carrier kinetics of perovskite thin films and presents a unified method for measuring carrier mobility and lifetime.

**Materials and Methods**

**Materials**

All chemicals were used as received, and the details are as follows: Methylammonium iodide (MAI, Greatcell Solar); Formamidinium iodide (FAI, Greatcell Solar); Cesium iodide (CsI, 99.9%, Alfa Aesar); Lead(II) iodide ($PbI_2$, 99.99%, TCI); Lead bromide ($PbBr_2$, 99.999%, Sigma-Aldrich); Cesium bromide (CsBr, 99.999%, Sigma-Aldrich); Anisole (99.7%, Sigma-Aldrich), *N,N*-dimethylformamide (DMF, 99.8%, Sigma-Aldrich); Dimethyl sulfoxide (DMSO, ≥99.9%; Sigma-Aldrich); Poly (methyl methacrylate) (PMMA, average $M_w$~120000 by GPC, Sigma-Aldrich).

**Sample preparation**

The typical procedure for preparation was generally consistent with previous work.(*34*) Specifically, quartz glass substrates (Corning, dimensions: 2.0 × 2.0 $cm^2$) were utilized in the study. These substrates were thoroughly cleaned using Seife Hellmanex III (2%, 50 °C) solution, acetone (20 °C), and isopropyl alcohol (20 °C) for 20 minutes. Afterwards, they underwent further cleaning using an oxygen plasma (Diener Zepto, 50 W, 13.56 MHz, 10 minutes). The solutions and films were prepared in a N$_2$-filled glovebox. The perovskite solution was prepared by mixing CsI (0.06 M), MAI (0.264 M), FAI (0.876 M), PbBr$_2$ (0.264 M), and PbI$_2$ (0.936 M). The mixture was stirred in a DMF:DMSO (3:1 volume ratio) solvent at 75°C until it was fully dissolved. ~0.06 mg/mL PMMA was added to the solution, which was then filtered through a PTFE filter (0.45 μm) prior to use. The precursor solution was spin-coated onto the substrates at 2000 rpm for 30 seconds (acceleration time = 3 seconds) and at 6000 rpm for 40 seconds (acceleration time = 5 seconds). The films were then treated with ~280 μl of anisole, which was dropped onto the film for 25 seconds prior to the end of the process. The films were finally annealed at 100°C for 20 minutes.

**Photoluminescence measurement**

Spectrally- and time-resolved resolved photoluminescence measurements were conducted using a gated CCD setup that comprised a pulsed UV-solid-state laser (dye laser, 343 nm wavelength, 100 Hz repetition rate), a spectrometer (SPEX 270M from Horiba Jobin Yvon), and an intensified CCD camera (iStar DH720 from Andor Solis). The applied excitation fluence, initial carrier concentration, and initial $\Delta E_F$ were approximately 1.79 μJ/cm$^2$, 5.61×10$^{16}$ cm$^{-3}$, and 1.44 eV, respectively. Additional information about the gated CCD setup can be found in ref. (*34*). Steady-state photoluminescence measurements were performed using a LuQY Pro setup (LP20-32, QYB Quantum Yield Berlin GmbH). Voltage-dependent photoluminescence measurements are carried out using a custom-made setup(*33, 71*) comprising a power supply, a Keithley 2400 SMU, a bias light source, and a CCD camera. The perovskite solar cells, with an active area of 3.0 × 3.0 mm$^2$, were illuminated with blue LED light (470 nm) of an intensity of 1 sun. During the measurement process, the CCD camera recorded the photoluminescence intensity at various voltages along with current-voltage (*J-V*) curves. Initially, the background was measured, followed by a flat-field correction to obtain the corrected PL intensity. The *J-V* curves were measured in both forward and reverse directions, with a voltage step of 0.02 V.

**Numerical Simulation**

Simulations were performed using self-developed MATLAB scripts. The detailed description is shown in Section 5 of the Supporting Information.

**References**
1. R. S. Crandall, Modeling of Thin-Film Solar-Cells - Nonuniform Field. *Journal of Applied Physics* **55**, 4418-4425 (1984).
2. R. S. Crandall, Transport in Hydrogenated Amorphous-Silicon P-I-N Solar-Cells. *Journal of Applied Physics* **53**, 3350-3352 (1982).
3. T. Kirchartz, J. Bisquert, I. Mora-Sero, G. Garcia-Belmonte, Classification of solar cells according to mechanisms of charge separation and charge collection. *Physical Chemistry Chemical Physics* **17**, 4007-4014 (2015).
4. D. J. Coutinho, G. C. Faria, D. T. Balogh, R. M. Faria, Influence of charge carriers mobility and lifetime on the performance of bulk heterojunction organic solar cells. *Solar Energy Materials and Solar Cells* **143**, 503-509 (2015).
5. J. G. Labram, E. E. Perry, N. R. Venkatesan, M. L. Chabinyc, Steady-state microwave conductivity reveals mobility-lifetime product in methylammonium lead iodide. *Applied Physics Letters* **113**, (2018).
6. I. Levine, S. Gupta, T. M. Brenner, D. Azulay, O. Millo, G. Hodes, D. Cahen, I. Balberg, Mobility–Lifetime Products in MAPbI3 Films. *The Journal of Physical Chemistry Letters* **7**, 5219-5226 (2016).
7. G. Dennler, A. J. Mozer, G. Juška, A. Pivrikas, R. Österbacka, A. Fuchsbauer, N. S. Sariciftci, Charge carrier mobility and lifetime versus composition of conjugated polymer/fullerene bulk-heterojunction solar cells. *Org Electron* **7**, 229-234 (2006).
8. J. Zhang, X. Liang, J. Min, J. Zhang, D. Zhang, C. Jin, S. Liang, P. Chen, L. Ling, J. Chen, Y. Shen, L. Wang, Effect of point defects trapping characteristics on mobility-lifetime (μτ) product in CdZnTe crystals. *J Cryst Growth* **519**, 41-45 (2019).
9. I. Levine, S. Gupta, A. Bera, D. Ceratti, G. Hodes, D. Cahen, D. Guo, T. J. Savenije, J. Ávila, H. J. Bolink, O. Millo, D. Azulay, I. Balberg, Can we use time-resolved measurements to get steady-state transport data for halide perovskites? *Journal of Applied Physics* **124**, (2018).
10. T. W. Crothers, R. L. Milot, J. B. Patel, E. S. Parrott, J. Schlipf, P. Müller-Buschbaum, M. B. Johnston, L. M. Herz, Photon Reabsorption Masks Intrinsic Bimolecular Charge-Carrier Recombination in CH3NH3PbI3 Perovskite. *Nano Lett* **17**, 5782-5789 (2017).
11. C. Cho, S. Feldmann, K. M. Yeom, Y.-W. Jang, S. Kahmann, J.-Y. Huang, T. C. J. Yang, M. N. T. Khayyat, Y.-R. Wu, M. Choi, J. H. Noh, S. D. Stranks, N. C. Greenham, Efficient vertical charge transport in polycrystalline halide perovskites revealed by four-dimensional tracking of charge carriers. *Nature Materials* **21**, 1388-1395 (2022).
12. S. Chattopadhyay, R. S. Kokenyesi, M. J. Hong, C. L. Watts, J. G. Labram, Resolving in-plane and out-of-plane mobility using time resolved microwave conductivity. *J Mater Chem C* **8**, 10761-10766 (2020).
13. R. Gegevičius, M. Franckevičius, V. Gulbinas, The Role of Grain Boundaries in Charge Carrier Dynamics in Polycrystalline Metal Halide Perovskites. *European Journal of Inorganic Chemistry* **2021**, 3519-3527 (2021).
14. D. H. Kim, J. Park, Z. Li, M. Yang, J.-S. Park, I. J. Park, J. Y. Kim, J. J. Berry, G. Rumbles, K. Zhu, 300% Enhancement of Carrier Mobility in Uniaxial-Oriented Perovskite Films Formed by Topotactic-Oriented Attachment. *Adv. Mater.* **29**, 1606831 (2017).

15. Q. Han, S.-H. Bae, P. Sun, Y.-T. Hsieh, Y. Yang, Y. S. Rim, H. Zhao, Q. Chen, W. Shi, G. Li, Y. Yang, Single Crystal Formamidinium Lead Iodide (FAPbI3): Insight into the Structural, Optical, and Electrical Properties. *Adv. Mater.* **28**, 2253-2258 (2016).
16. A. A. Zhumekenov, M. I. Saidaminov, M. A. Haque, E. Alarousu, S. P. Sarmah, B. Murali, I. Dursun, X.-H. Miao, A. L. Abdelhady, T. Wu, O. F. Mohammed, O. M. Bakr, Formamidinium Lead Halide Perovskite Crystals with Unprecedented Long Carrier Dynamics and Diffusion Length. *ACS Energy Lett.* **1**, 32-37 (2016).
17. W. Rehman, R. L. Milot, G. E. Eperon, C. Wehrenfennig, J. L. Boland, H. J. Snaith, M. B. Johnston, L. M. Herz, Charge-Carrier Dynamics and Mobilities in Formamidinium Lead Mixed-Halide Perovskites. *Adv. Mater.* **27**, 7938-7944 (2015).
18. G. E. Eperon, S. D. Stranks, C. Menelaou, M. B. Johnston, L. M. Herz, H. J. Snaith, Formamidinium lead trihalide: a broadly tunable perovskite for efficient planar heterojunction solar cells. *Energ Environ Sci* **7**, 982-988 (2014).
19. S. D. Stranks, G. E. Eperon, G. Grancini, C. Menelaou, M. J. P. Alcocer, T. Leijtens, L. M. Herz, A. Petrozza, H. J. Snaith, Electron-Hole Diffusion Lengths Exceeding 1 Micrometer in an Organometal Trihalide Perovskite Absorber. *Science* **342**, 341-344 (2013).
20. G. Xing, N. Mathews, S. Sun, S. S. Lim, Y. M. Lam, M. Gratzel, S. Mhaisalkar, T. C. Sum, Long-range balanced electron- and hole-transport lengths in organic-inorganic CH3NH3PbI3. *Science* **342**, 344-347 (2013).
21. Z. Guo, J. S. Manser, Y. Wan, P. V. Kamat, L. Huang, Spatial and temporal imaging of long-range charge transport in perovskite thin films by ultrafast microscopy. *Nature Communications* **6**, 7471 (2015).
22. R. L. Milot, G. E. Eperon, H. J. Snaith, M. B. Johnston, L. M. Herz, Temperature-Dependent Charge-Carrier Dynamics in CH3NH3PbI3 Perovskite Thin Films. *Adv. Funct. Mater.* **25**, 6218-6227 (2015).
23. C. S. Ponseca, T. J. Savenije, M. Abdellah, K. Zheng, A. Yartsev, T. Pascher, T. Harlang, P. Chabera, T. Pullerits, A. Stepanov, J.-P. Wolf, V. Sundström, Organometal Halide Perovskite Solar Cell Materials Rationalized: Ultrafast Charge Generation, High and Microsecond-Long Balanced Mobilities, and Slow Recombination. *Journal of the American Chemical Society* **136**, 5189-5192 (2014).
24. C. La-o-vorakiat, T. Salim, J. Kadro, M.-T. Khuc, R. Haselsberger, L. Cheng, H. Xia, G. G. Gurzadyan, H. Su, Y. M. Lam, R. A. Marcus, M.-E. Michel-Beyerle, E. E. M. Chia, Elucidating the role of disorder and free-carrier recombination kinetics in CH3NH3PbI3 perovskite films. *Nature Communications* **6**, 7903 (2015).
25. E. M. Hutter, G. E. Eperon, S. D. Stranks, T. J. Savenije, Charge Carriers in Planar and Meso-Structured Organic–Inorganic Perovskites: Mobilities, Lifetimes, and Concentrations of Trap States. *The Journal of Physical Chemistry Letters* **6**, 3082-3090 (2015).
26. O. G. Reid, M. Yang, N. Kopidakis, K. Zhu, G. Rumbles, Grain-Size-Limited Mobility in Methylammonium Lead Iodide Perovskite Thin Films. *ACS Energy Lett.* **1**, 561-565 (2016).
27. D. A. Valverde-Chávez, C. S. Ponseca, C. C. Stoumpos, A. Yartsev, M. G. Kanatzidis, V. Sundström, D. G. Cooke, Intrinsic femtosecond charge generation dynamics in single crystal CH3NH3PbI3. *Energ Environ Sci* **8**, 3700-3707 (2015).

28. O. E. Semonin, G. A. Elbaz, D. B. Straus, T. D. Hull, D. W. Paley, A. M. van der Zande, J. C. Hone, I. Kymissis, C. R. Kagan, X. Roy, J. S. Owen, Limits of Carrier Diffusion in n-Type and p-Type CH3NH3PbI3 Perovskite Single Crystals. *The Journal of Physical Chemistry Letters* **7**, 3510-3518 (2016).
29. Q. Dong, Y. Fang, Y. Shao, P. Mulligan, J. Qiu, L. Cao, J. Huang, Solar cells. Electron-hole diffusion lengths > 175 mum in solution-grown CH3NH3PbI3 single crystals. *Science* **347**, 967-970 (2015).
30. M. I. Saidaminov, A. L. Abdelhady, B. Murali, E. Alarousu, V. M. Burlakov, W. Peng, I. Dursun, L. Wang, Y. He, G. Maculan, A. Goriely, T. Wu, O. F. Mohammed, O. M. Bakr, High-quality bulk hybrid perovskite single crystals within minutes by inverse temperature crystallization. *Nature Communications* **6**, 7586 (2015).
31. D. Shi, V. Adinolfi, R. Comin, M. Yuan, E. Alarousu, A. Buin, Y. Chen, S. Hoogland, A. Rothenberger, K. Katsiev, Y. Losovyj, X. Zhang, P. A. Dowben, O. F. Mohammed, E. H. Sargent, O. M. Bakr, Low trap-state density and long carrier diffusion in organolead trihalide perovskite single crystals. *Science* **347**, 519-522 (2015).
32. O. Gunawan, S. R. Pae, D. M. Bishop, Y. Virgus, J. H. Noh, N. J. Jeon, Y. S. Lee, X. Shao, T. Todorov, D. B. Mitzi, B. Shin, Carrier-resolved photo-Hall effect. *Nature* **575**, 151-155 (2019).
33. J. Siekmann, A. Kulkarni, S. Akel, B. Klingebiel, M. Saliba, U. Rau, T. Kirchartz, Characterizing the Influence of Charge Extraction Layers on the Performance of Triple-Cation Perovskite Solar Cells. *Adv Energy Mater* **13**, 2300448 (2023).
34. Y. Yuan, G. Yan, C. Dreessen, T. Rudolph, M. Hülsbeck, B. Klingebiel, J. Ye, U. Rau, T. Kirchartz, Shallow defects and variable photoluminescence decay times up to 280 µs in triple-cation perovskites. *Nature Materials* **23**, 391 - 397 (2024).
35. D. Kiermasch, A. Baumann, M. Fischer, V. Dyakonov, K. Tvingstedt, Revisiting lifetimes from transient electrical characterization of thin film solar cells; a capacitive concern evaluated for silicon, organic and perovskite devices. *Energy & Environmental Science* **11**, 629-640 (2018).
36. L. Krückemeier, Z. Liu, B. Krogmeier, U. Rau, T. Kirchartz, Consistent Interpretation of Electrical and Optical Transients in Halide Perovskite Layers and Solar Cells. *Advanced Energy Materials* **11**, 2102290 (2021).
37. L. Krückemeier, Z. Liu, T. Kirchartz, U. Rau, Quantifying Charge Extraction and Recombination Using the Rise and Decay of the Transient Photovoltage of Perovskite Solar Cells. *Adv. Mater.* **35**, 2300872 (2023).
38. L. Castaner, E. Vilamajo, J. Llaberia, J. Garrido, Investigations of the OCVD transients in solar cells. *Journal of Physics D: Applied Physics* **14**, 1867 (1981).
39. A. Cuevas, F. Recart, Capacitive effects in quasi-steady-state voltage and lifetime measurements of silicon devices. *Journal of Applied Physics* **98**, (2005).
40. J. Ye, M. M. Byranvand, C. O. Martínez, R. L. Z. Hoye, M. Saliba, L. Polavarapu, Defect Passivation in Lead-Halide Perovskite Nanocrystals and Thin Films: Toward Efficient LEDs and Solar Cells. *Angewandte Chemie International Edition* **60**, 21636-21660 (2021).
41. A. R. Srimath Kandada, A. Petrozza, Research Update: Luminescence in lead halide perovskites. *APL Materials* **4**, (2016).

42. V. S. Chirvony, S. González-Carrero, I. Suárez, R. E. Galian, M. Sessolo, H. J. Bolink, J. P. Martínez-Pastor, J. Pérez-Prieto, Delayed Luminescence in Lead Halide Perovskite Nanocrystals. *The Journal of Physical Chemistry C* **121**, 13381-13390 (2017).
43. D. N. Dirin, L. Protesescu, D. Trummer, I. V. Kochetygov, S. Yakunin, F. Krumeich, N. P. Stadie, M. V. Kovalenko, Harnessing Defect-Tolerance at the Nanoscale: Highly Luminescent Lead Halide Perovskite Nanocrystals in Mesoporous Silica Matrixes. *Nano Lett* **16**, 5866-5874 (2016).
44. H. Jin, E. Debroye, M. Keshavarz, I. G. Scheblykin, M. B. J. Roeffaers, J. Hofkens, J. A. Steele, It's a trap! On the nature of localised states and charge trapping in lead halide perovskites. *Materials Horizons* **7**, 397-410 (2020).
45. F. Deschler, M. Price, S. Pathak, L. E. Klintberg, D.-D. Jarausch, R. Higler, S. Hüttner, T. Leijtens, S. D. Stranks, H. J. Snaith, M. Atatüre, R. T. Phillips, R. H. Friend, High Photoluminescence Efficiency and Optically Pumped Lasing in Solution-Processed Mixed Halide Perovskite Semiconductors. *The Journal of Physical Chemistry Letters* **5**, 1421-1426 (2014).
46. R. K. Ahrenkiel, Minority-Carrier Lifetime in III-V Semiconductors. *Minority Carriers in Iii-V Semiconductors: Physics and Applications* **39**, 39-150 (1993).
47. C. J. Hages, A. Redinger, S. Levcenko, H. Hempel, M. J. Koeper, R. Agrawal, D. Greiner, C. A. Kaufmann, T. Unold, Identifying the Real Minority Carrier Lifetime in Nonideal Semiconductors: A Case Study of Kesterite Materials. *Adv Energy Mater* **7**, 1700167 (2017).
48. T. Unold, L. Gutay, "Photoluminescence Analysis of Thin-Film Solar Cells" in *Advanced Characterization Techniques for Thin Film Solar Cells* (Wiley-VCH Verlag GmbH & Co. KGaA, 2011), pp. 151-175.
49. G. D. Gilliland, Photoluminescence spectroscopy of crystalline semiconductors. *Materials Science and Engineering: R: Reports* **18**, 99-399 (1997).
50. C. Huang, S. Wu, A. M. Sanchez, J. J. P. Peters, R. Beanland, J. S. Ross, P. Rivera, W. Yao, D. H. Cobden, X. Xu, Lateral heterojunctions within monolayer MoSe2–WSe2 semiconductors. *Nature Materials* **13**, 1096-1101 (2014).
51. M. A. Reshchikov, Measurement and analysis of photoluminescence in GaN. *Journal of Applied Physics* **129**,  (2021).
52. H. Nan, Z. Wang, W. Wang, Z. Liang, Y. Lu, Q. Chen, D. He, P. Tan, F. Miao, X. Wang, J. Wang, Z. Ni, Strong Photoluminescence Enhancement of MoS2 through Defect Engineering and Oxygen Bonding. *ACS Nano* **8**, 5738-5745 (2014).
53. X. Zhang, H. Dong, W. Hu, Organic Semiconductor Single Crystals for Electronics and Photonics. *Adv. Mater.* **30**, 1801048 (2018).
54. H. Yu, X. Cui, X. Xu, W. Yao, Valley excitons in two-dimensional semiconductors. *National Science Review* **2**, 57-70 (2015).
55. V. Sarritzu, N. Sestu, D. Marongiu, X. Chang, S. Masi, A. Rizzo, S. Colella, F. Quochi, M. Saba, A. Mura, G. Bongiovanni, Optical determination of Shockley-Read-Hall and interface recombination currents in hybrid perovskites. *Scientific Reports* **7**, 44629 (2017).
56. M. Stolterfoht, C. M. Wolff, J. A. Márquez, S. S. Zhang, C. J. Hages, D. Rothhardt, S. Albrecht, P. L. Burn, P. Meredith, T. Unold, D. Neher, Visualization and suppression of interfacial recombination for high-efficiency large-area pin perovskite solar cells. *Nat. Energy* **3**, 847-854 (2018).

57. S. D. Stranks, V. M. Burlakov, T. Leijtens, J. M. Ball, A. Goriely, H. J. Snaith, Recombination Kinetics in Organic-Inorganic Perovskites: Excitons, Free Charge, and Subgap States. *Physical Review Applied* **2**, 034007 (2014).
58. M. Stolterfoht, V. M. Le Corre, M. Feuerstein, P. Caprioglio, L. J. A. Koster, D. Neher, Voltage-Dependent Photoluminescence and How It Correlates with the Fill Factor and Open-Circuit Voltage in Perovskite Solar Cells. *ACS Energy Lett.* **4**, 2887-2892 (2019).
59. Y. Guo, O. Yaffe, T. D. Hull, J. S. Owen, D. R. Reichman, L. E. Brus, Dynamic emission Stokes shift and liquid-like dielectric solvation of band edge carriers in lead-halide perovskites. *Nature Communications* **10**, 1175 (2019).
60. A. D. Wright, C. Verdi, R. L. Milot, G. E. Eperon, M. A. Pérez-Osorio, H. J. Snaith, F. Giustino, M. B. Johnston, L. M. Herz, Electron–phonon coupling in hybrid lead halide perovskites. *Nature Communications* **7**, 11755 (2016).
61. P. Caprioglio, M. Stolterfoht, C. M. Wolff, T. Unold, B. Rech, S. Albrecht, D. Neher, On the Relation between the Open-Circuit Voltage and Quasi-Fermi Level Splitting in Efficient Perovskite Solar Cells. *Adv Energy Mater* **9**, 1901631 (2019).
62. F. Peña-Camargo, J. Thiesbrummel, H. Hempel, A. Musiienko, V. M. Le Corre, J. Diekmann, J. Warby, T. Unold, F. Lang, D. Neher, M. Stolterfoht, Revealing the doping density in perovskite solar cells and its impact on device performance. *Applied Physics Reviews* **9**, 021409 (2022).
63. H. Hempel, M. Stolterfoht, O. Karalis, T. Unold, The Potential of Geminate Pairs in Lead Halide Perovskite revealed via Time-resolved Photoluminescence. *arXiv*, 2409.06382 (2024).
64. P. Würfel, T. Trupke, T. Puzzer, E. Schaffer, W. Warta, S. W. Glunz, Diffusion lengths of silicon solar cells from luminescence images. *Journal of Applied Physics* **101**, 123110 (2007).
65. A. Roigé, J. Alvarez, A. Jaffré, T. Desrues, D. Muñoz, I. Martín, R. Alcubilla, J.-P. Kleider, Effects of photon reabsorption phenomena in confocal micro-photoluminescence measurements in crystalline silicon. *Journal of Applied Physics* **121**, (2017).
66. F. Staub, I. Anusca, D. C. Lupascu, U. Rau, T. Kirchartz, Effect of reabsorption and photon recycling on photoluminescence spectra and transients in lead-halide perovskite crystals. *J Phys-Mater* **3**, (2020).
67. L. Krückemeier, B. Krogmeier, Z. Liu, U. Rau, T. Kirchartz, Understanding Transient Photoluminescence in Halide Perovskite Layer Stacks and Solar Cells. *Adv Energy Mater* **11**, 2003489 (2021).
68. P. Wurfel, The Chemical-Potential of Radiation. *J Phys C Solid State* **15**, 3967-3985 (1982).
69. O. J. Sandberg, J. Kurpiers, M. Stolterfoht, D. Neher, P. Meredith, S. Shoaee, A. Armin, On the Question of the Need for a Built-In Potential in Perovskite Solar Cells. *Adv. Mater. Interfaces* **7**, 2000041 (2020).
70. U. Rau, V. Huhn, B. E. Pieters, Luminescence Analysis of Charge-Carrier Separation and Internal Series-Resistance Losses in Cu(In,Ga)Se2 Solar Cells. *Physical Review Applied* **14**, 014046 (2020).
71. S. Akel, A. Kulkarni, U. Rau, T. Kirchartz, Relevance of Long Diffusion Lengths for Efficient Halide Perovskite Solar Cells. *PRX Energy* **2**, 013004 (2023).

72. K. Misiakos, F. A. Lindholm, Minority-carrier accumulation at the base edge of a junction space-charge region under short-circuit conditions. *Solid-State Electronics* **30**, 755-758 (1987).

**Acknowledgments**

**Funding:** We acknowledge funding by
  Helmholtz Association via the POF IV funding
  Helmholtz Association via the innovation platform "SolarTAP – A Solar Technology Acceleration Platform"
  Helmholtz Association via the project "Beschleunigter Transfer der nächsten Generation von Solarzellen in die Massenfertigung - Zukunftstechnologie Tandem-Solarzellen"
  Helmholtz Association via the Helmholtz.AI project AISPA - AI-driven instantaneous solar cell property analysis
  the Deutsche Forschungsgemeinschaft (German Research Foundation) via the project "Correlating Defect Densities with Recombination Losses in Halide-Perovskite Solar Cells"

**Data and materials availability:** All data are available in the main text or the supplementary materials. The data presented in the main text and the simulation codes are available at DOI: 10.5281/zenodo.13757008.